\documentclass[sigconf,natbib=false]{acmart} % ,anonymous

\AtBeginDocument{%
  }

\copyrightyear{2024}
\acmYear{2024}
\setcopyright{rightsretained}
\acmConference[ARES 2024]{The 19th International Conference on Availability, Reliability and Security}{July 30-August 2, 2024}{Vienna, Austria}
\acmBooktitle{The 19th International Conference on Availability, Reliability and Security (ARES 2024), July 30-August 2, 2024, Vienna, Austria}\acmDOI{10.1145/3664476.3669935}
\acmISBN{979-8-4007-1718-5/24/07}

\RequirePackage[
  datamodel=acmdatamodel,
  style=acmnumeric,
  ]{biblatex}
\usepackage{xcolor}
\usepackage{color, colortbl}
\usepackage[acronym]{glossaries}
\newacronym{asn1}{ASN.1}{Abstract Syntax Notation One}
\newacronym{ber}{BER}{Basic Encoding Rules}
\newacronym{ca}{CA}{Certificate Authority}
\newacronym{cer}{CER}{Canonical Encoding Rules}
\newacronym{cve}{CVE}{Common Vulnerabilities and Exposures}
\newacronym{cwe}{CWE}{Common Weakness Enumeration}
\newacronym{der}{DER}{Distinguished Encoding Rules}
\newacronym{foss}{FOSS}{Free and open-source software}
\newacronym{iec}{IEC}{International Electrotechnical Commission}
\newacronym{ietf}{IETF}{Internet Engineering Task Force}
\newacronym{iso}{ISO}{International Organization for Standardization}
\newacronym{itu-t}{ITU-T}{ITU Telecommunication Standardization Sector}
\newacronym{itu}{ITU}{International Telecommunication Union}
\newacronym{pki}{PKI}{Public Key Infrastructure}
\newacronym{png}{PNG}{Portable Network Graphics}
\newacronym{ra}{RA}{Registration Authority}
\newacronym{rp}{RP}{Relying Party}
\newacronym{smime}{S/MIME}{Secure/Multipurpose Internet Mail Extensions}
\newacronym{tls}{TLS}{Transport Layer Security}
\newacronym{un}{UN}{United Nations}
\newacronym{url}{URL}{Uniform Resource Locator}
\newacronym{json}{JSON}{JavaScript Object Notation}
\newacronym{oid}{OID}{Object Identifier}
\newacronym{iqr}{IQR}{Inner Quartile Range}
\newacronym{api}{API}{Application Programming Interface}
\newacronym{abi}{ABI}{Application Binary Interface}
\newacronym{zst}{zstd}{Zstandard}
\newacronym{ecdsa}{ECDSA}{Elliptic Curve Digital Signature Algorithm}
\newacronym{nss}{NSS}{Network Security Services}
\newacronym{uri}{URI}{Uniform Resource Identifier}

\usepackage{subcaption}
\usepackage{enumitem}
\usepackage{fancyvrb}
\usepackage{blindtext}
\usepackage{arydshln}
\usepackage[shortcuts]{extdash}

\newcommand{\resQuestion}{“Are there user-visible and security-related differences between X.509 certificate parsers?”}
\newcommand{\exampleError}[1]{An example error string for this category is: “#1”}
\newcommand*\rot{\rotatebox{75}}

\begin{document}

\title{ParsEval: Evaluation of Parsing Behavior using Real-world Out-in-the-wild X.509 Certificates}

%%
%% The "author" command and its associated commands are used to define
%% the authors and their affiliations.
%% Of note is the shared affiliation of the first two authors, and the
%% "authornote" and "authornotemark" commands
%% used to denote shared contribution to the research.

% https://tex.stackexchange.com/a/505807
\author{Stefan Tatschner}
% \authornote{Both authors contributed equally to this research.}
\orcid{0000-0002-2288-9010}
\affiliation{%
  \institution{Fraunhofer Institute AISEC}
  \streetaddress{Lichtenbergstraße 11}
  \city{Garching bei München}
  \country{Germany}
  \postcode{85748}
}
\affiliation{%
  \institution{University of Limerick}
  \city{Limerick}
  \country{Ireland}
}
\email{stefan.tatschner@aisec.fraunhofer.de}

\author{Sebastian N. Peters}
% \authornotemark[1]
\orcid{0009-0007-6421-4023}
\affiliation{%
  \institution{Technical University of Munich} % TUM CIT, Dept. of Comp. Eng., 
  \city{Garching bei München}
  \country{Germany}
}
\affiliation{%
  \institution{Fraunhofer Institute AISEC}
  \streetaddress{Lichtenbergstraße 11}
  \city{Garching bei München}
  \country{Germany}
  \postcode{85748}
}
\email{sebastian.peters@aisec.fraunhofer.de}

\author{Michael P. Heinl}
% \authornotemark[1]
\orcid{0000-0002-1094-4828}
\affiliation{%
  \institution{Technical University of Munich} % TUM CIT, Dept. of Comp. Eng., 
  \city{Garching bei München}
  \country{Germany}
}
\affiliation{%
  \institution{Fraunhofer Institute AISEC}
  \streetaddress{Lichtenbergstraße 11}
  \city{Garching bei München}
  \country{Germany}
  \postcode{85748}
}
\email{michael.heinl@aisec.fraunhofer.de}

\author{Tobias Specht}
\orcid{0009-0001-7615-7579}
\affiliation{%
  \institution{Fraunhofer Institute AISEC}
  \streetaddress{Lichtenbergstraße 11}
  \city{Garching bei München}
  \country{Germany}
  \postcode{85748}
}
\email{tobias.specht@aisec.fraunhofer.de}

% Department of Electronic and Computer Engineering, University of Limerick, Ireland; 
\author{Thomas Newe}
\orcid{0000-0002-3375-8200}
\affiliation{%
  \institution{University of Limerick}
  \city{Limerick}
  \country{Ireland}
}
\email{thomas.newe@ul.ie}

\begin{abstract}
X.509 certificates play a crucial role in establishing secure communication over the internet by enabling authentication and data integrity.
Equipped with a rich feature set, the X.509 standard is defined by multiple, comprehensive ISO/IEC documents.
Due to its internet-wide usage, there are different implementations in multiple programming languages leading to a large and fragmented ecosystem.
This work addresses the research question \resQuestion{}.
Relevant libraries offering APIs for parsing X.509 certificates were investigated and an appropriate test suite was developed.
From 34 libraries 6 were chosen for further analysis.
The X.509 parsing modules of the chosen libraries were called with 186,576,846 different certificates from a real-world dataset and the observed error codes were investigated.
This study reveals an anomaly in wolfSSL's X.509 parsing module and that there are fundamental differences in the ecosystem.
While related studies nowadays mostly focus on fuzzing techniques resulting in artificial certificates, this study confirms that available X.509 parsing modules differ largely and yield different results, even for real-world out-in-the-wild certificates.
\end{abstract}

\ccsdesc[500]{Security and privacy~Systems security}
\ccsdesc[500]{Security and privacy~Security services}
\ccsdesc[500]{Security and privacy~Software security engineering}
\ccsdesc[300]{Security and privacy~Web application security}
\ccsdesc[300]{Security and privacy~Security protocols}
\ccsdesc[300]{Security and privacy~Web protocol security}
\ccsdesc[300]{Security and privacy~Authentication}
\ccsdesc[500]{Software and its engineering~Software verification and validation}

%%
%% Keywords. The author(s) should pick words that accurately describe
%% the work being presented. Separate the keywords with commas.
\keywords{digital certificates, X.509, ASN.1, parsing, TLS libraries, conformity testing}
%% A "teaser" image appears between the author and affiliation
%% information and the body of the document, and typically spans the
%% page.
% \begin{teaserfigure}
%   \includegraphics[width=\textwidth]{samples/sampleteaser}
%   \caption{Seattle Mariners at Spring Training, 2010.}
%   \Description{Enjoying the baseball game from the third-base
%   seats. Ichiro Suzuki preparing to bat.}
%   \label{fig:teaser}
% \end{teaserfigure}

%\received{30 April 2023}
%\received[revised]{30 April 2023}
%\received[accepted]{30 April 2023}

%%
%% This command processes the author and affiliation and title
%% information and builds the first part of the formatted document.
\maketitle

\section{Introduction}

\small
\begin{table*}
\caption{Examples of Recent Vulnerabilities in \gls{asn1} or X.509 Components from 2020 -- 2023.}
\centering
    \begin{tabular}{llllll}
  \toprule
CVE & CWE & Component & Affected Project & Severity \\
  \midrule
%  CVE-2024-22039 & CWE-120: Buffer Copy w/o Checking Size  of Input & X.509 &  Siemens Fire Panels  & Critical\\ 
% Hinweis von Nikolai
% Aber passt doch nicht so ganz hier rein...
%  \hdashline
  CVE-2023-0464 & CWE-295: Improper Certificate Validation & X.509 &  OpenSSL  & High\\ 
  \hdashline
  CVE-2023-2650 & CWE-770: Allocation of Resources Without Limits or Throttling & \gls{asn1} &  OpenSSL  & Medium\\ 
  \hdashline
  CVE-2023-0286 & CWE-843: Access of Resource Using Incompatible Type & \gls{asn1} &  OpenSSL  & High\\ 
  \hdashline
  CVE-2022-0778 & CWE-835: Loop with Unreachable Exit Condition & X.509 &  OpenSSL  & High\\
  \hdashline
  CVE-2022-4203 & CWE-125: Out-of-bounds Read & X.509 & OpenSSL & Critical\\
  \hdashline
  CVE-2021-34558 & CWE-295: Improper Certificate Validation & X.509 & Go stdlib & Medium\\
  \hdashline
  CVE-2021-3712 & CWE-125: Out-of-bounds Read & \gls{asn1} & OpenSSL & High\\
  \hdashline
  CVE-2021-46848 & CWE-125: Out-of-bounds Read & \gls{asn1} & GnuTLS & Critical\\
  \hdashline
  CVE-2020-7919 & CWE-295: Improper Certificate Validation & X.509 & Go stdlib & High\\
  \hdashline
  CVE-2020-36477 & CWE-295: ImproperCertificate Validation & X.509 & Mbed~TLS & Medium\\
  \bottomrule
    \end{tabular}
    \label{tab:cve-overview}
\end{table*}
\normalsize

% Why are X.509 certificates important?
Secure communication utilizing authentication and data integrity is an important requirement for today's internet.
X.509 certificates are therefore a crucial building block of the relevant software stack.
They are defined by the \gls{iso} and \gls{iec} in ISO/IEC 9594-8~\cite{iso-iec-9594-8}.
However, this version defined by ISO/IEC is broad in its applicability.
Therefore, a suitable profile meeting the requirements of internet use was necessary.
In 2008, the \gls{ietf} published RFC~5280~\cite{rfc5280} that clarifies these requirements and defines a corresponding \gls{pki} for internet, electronic mail, and IPsec applications. Moreover, there is also an increased utilization of \gls{pki} to secure interconnected industrial environments~\cite{heinl2023} which renders digital certificates even more critical.
The \gls{pki} paradigm includes different components~\cite{rfc5280}, such as \gls{ca}, \gls{ra}, and \gls{rp}, providing a framework for managing digital certificates.
After initial registration of its owner, subscriber, or subject towards an \gls{ra}, these certificates are signed by a \gls{ca} and used by an \gls{rp} to verify the authenticity of subjects such as individuals, servers, or organizations in online communications.

% Why do we not just trust what is already there?
Previous research~\cite{10.1145/3338466.3358917} suggests assessing a given \gls{ca}'s trustworthiness before relying on digital certificates issued by a specific \gls{ca}.
However, questions regarding trustworthiness and reliability cannot only arise on the side of the \gls{ca}, but also regarding the software stack used by the \gls{rp} as demonstrated by various vulnerabilities listed in \autoref{tab:cve-overview}.
Many of them are related to erroneous or inconsistent parsing of the underlying X.509 data structures.
To understand the reasons, the multi-faceted nature of these data structures needs to be investigated.

% Background to understand the actual problem
The X.509 certificate format defines a standardized container structure for exchanging public key information.
A typical X.509 certificate contains various fields, for instance the subject's identity (e.g., a website's domain name), the issuer's identity (the \gls{ca} that issued the certificate), the associated public key, and further relevant metadata.
Over time, X.509 has become widely adopted as the de facto standard for digital certificates.
It is used in various critical applications, including \gls{smime} for email, industrial machine-to-machine communication~\cite{heinl2023}, or \gls{tls} for general purpose internet connections.
X.509 relies on the \gls{asn1} standard which was initially developed in 1984 and published as X.680~\cite{x680}.
\gls{asn1} specifies a flexible language for describing structured data.
Consequently, \gls{asn1} can be used by computer systems to communicate with each other, regardless of their underlying hardware or programming language.
Several especially networking-related standards rely on \gls{asn1} for defining the structure and encoding rules of data types;
\gls{asn1} provides a set of rules for defining data types and their encodings.
It defines a wide range of primitive data types such as integers, strings, dates, and booleans, as well as complex types like sequences, sets, and choice types.
These data types can be composed and nested to represent more complex structures.

% X.509's and \gls{asn1}'s multipurpose character as the root of evil and what to do against it.
Different encoding rules exist, such as \gls{ber}, \gls{cer}, or \gls{der}.
Each rule has its own advantages and trade-offs in terms of efficiency, complexity, and interoperability.
\gls{asn1} plays a crucial role in ensuring interoperability between different systems by providing a standardized way to describe and encode data structures.
It has been widely used in industries like telecommunications, security, finance, and healthcare, where reliable data exchange and protocol compatibility are essential.

Overall, \gls{asn1} and X.509 are used for a countless number of security-related use cases on a plethora of different platforms using a variety of programming languages.
This heterogeneity leads to the fact that different parsers lead to different results although they should be consistent.
This deviant behavior can constitute an attack surface which might be overlooked.
To investigate this hypothesis, this paper addresses the following research question \resQuestion{}.

%\begin{enumerate}[label=\textbf{RQ\arabic*:}, leftmargin=*]
%   \item Which differences between X.509 certificate parsers exist?
%   \item What are the security implications of these differences?
%   \item Can certificates which produce abnormal parsing behavior be mapped to specific CAs?
%\end{enumerate}

\section{Related Work}

In 2014, Brubaker~et~al. published the paper “Using Frankencerts for Automated Adversarial Testing of Certificate Validation in SSL/TLS Implementations”~\cite{frankencert2014} in which they developed a methodology for large-scale testing of certificate validation logic in \gls{tls} implementations.
The authors used a differential testing-based approach by creating syntactically valid certificates containing false metadata.
Those certificates were called “frankencerts”.
This paper uncovered 208 discrepancies between popular \gls{tls} implementations.
Their approach was refined in 2015 by Chen et al. in “Guided differential testing of certificate validation in SSL/TLS implementations”~\cite{10.1145/2786805.2786835} informally called “mucert”.
A further refinement is “transcert” by Zhu et al. which was published in 2020 as “Guided, Deep Testing of X.509 Certificate Validation via Coverage Transfer Graphs”~\cite{9240633}.

In 2015, Kaloper-Meršinjak~et~al. published the paper “Not-Quite-So-Broken TLS: Lessons in Re-Engineering a security protocol specification and implementation”~\cite{190896}.
The authors developed a full \gls{tls} implementation using a novel approach to security protocol specification and implementation.
Their declarative programming approach provides both, the specification and the implementation of the protocol at the same time.
The authors' main point of criticism against the \gls{asn1} and X.509 standards is the high complexity of these standards.
Using the developed approach, the X.509 validation logic covered in $\approx 7,000$ lines of prose text could be reduced to $\approx 300$ lines of code.
This work shows that technical specifications written in prose are highly redundant.
This redundancy could increase the number of human errors during the implementation.

In 2023, Sorniotti et al. published the paper “Go or No Go: Differential Fuzzing of Native and C Libraries”~\cite{sorniotti2023go}.
The authors identified differences in parsing between several libraries implementing core functionality, such as \gls{png} encoding/decoding or file compression.
Native Go variants of said libraries were compared with mature C/C++ variants using a differential fuzzing based approach.
One of their findings turned out to be a security problem which was disclosed to Google\footnote{\url{https://go-review.googlesource.com/c/go/+/395734/}} and fixed in Go version 1.19.
The paper concludes that despite Go provides strong memory safety guarantees, subtle parsing differences may still negatively impact the security of native Go libraries.

In 2023, Chen et al. published the paper “SBDT: Search-Based Differential Testing of Certificate Parsers in SSL/TLS Implementations”~\cite{10.1145/3597926.3598110}.
The authors present a novel approach for generating certificates that can be used to test specific parts of the certificate parsing modules.
The authors present experimental results that show that their approach and its prototype tool are effective and efficient in finding bugs in certificate parsers of SSL/TLS implementations.

\small

\begin{table*}
\caption{Overview of identified production\=/ready libraries offering certificate parsing modules, sorted by Github stars.}
\centering
    \begin{tabular}{lcllll}
  \toprule
Project & $\star$ & Backed by & Language & References\\
  \midrule
  Go stdlib & 115k & Google & Go, Assembly & Google Services\\
  \hdashline
  OpenSSL & 22.7k & OpenSSL project & C, Assembly & Apache, nginx, OpenVPN, “it's everywhere”\\
  \hdashline
  python-cryptography & 5.8k & Python community & Python, Rust & certbot, python-telegram-bot, aiohttp, oauthlib, docker-py\\
  \hdashline
  Mbed~TLS & 4.4k & trustedfirmware.org, ARM & C, Assembly  & embedded projects, OpenVPN, Hiawatha Webserver, PowerDNS\\
  \hdashline
  wolfSSL & 2k & wolfSSL Inc. & C, Assembly & embedded projects, OpenSSL API compatible \\
  \hdashline
  GnuTLS & n.a. & Free Software Foundation & C, Assembly & GNOME, Exim, Wireshark, CUPS, Synology DiskStation Manager\\
  \bottomrule
    \end{tabular}
    \label{tab:lib-overview}
\end{table*}

\normalsize

In 2023, Tatschner et al. conducted a literature research on current implementations of the QUIC protocol \cite{tatschner2023ares}.
They created an overview of relevant libraries that implement the QUIC protocol.
Tatschner et al. also published an extended version of their overview\footnote{\url{https://rumpelsepp.org/projects/quic-overview}} providing information about the used \gls{tls} libraries.
Since the paper examined and documented the relevance of these \gls{tls} libraries, its choice of libraries was considered for \autoref{tab:lib-overview}.

\section{Methodology}

This research consists of multiple, consecutive working steps.
From a conceptional point of view, the methodology is structured as follows:

\begin{enumerate}
    \item \textbf{Choice of libraries}: 
    The number of libraries for this analysis was limited by the following constraints, similar as in Tatschner et al. \cite{tatschner2023ares}:
    \begin{enumerate}
        \item Only implementations that consider themselves as production\=/ready or implementations that are already used by popular applications, such as curl\footnote{\url{https://curl.se}}. 
        \item Only \gls{foss} projects, because an important point are publicly available resources, such as documentation for developers.
        \item \gls{api} support for X.509 certificate parsing without triggering the validation of the trust chain.
    \end{enumerate}
    
    \item \textbf{Test data}:
    Appropriate, real-world test data was identified and collected.
    During the research, it became apparent that creating an own crawler similar to the one in \cite{frankencert2014} was infeasible, due to the large amount of data required for meaningful results.
    Therefore, an alternative approach was used:
    A ready-to-use 11.6~TiB dataset consisting of collected X.509 certificates made available for public research by Censys~\cite{censys2015}\footnote{\url{https://censys.io}} was used after converting it into an appropriate format.
    \item \textbf{Data acquisition}: 
    A random set of 2,000 batch files consisting of 186,576,846 unique certificates was picked from the downloaded test data corpus.
    The tests were performed with these certificates and all the results were collected in a database for further inspection.
    In contrast to the investigated related work, e.g., frankencert or transcert, only unmodified real-world certificates were used for this study.
    \item \textbf{Data evaluation and interpretation}: The collected test results were evaluated with statistical methods and logical reasoning.
\end{enumerate}

\subsection{Selection of Libraries}

\autoref{tab:lib-overview} shows the considered libraries for this paper.
The libraries were chosen with a similar approach as presented in \cite{tatschner2023ares}.
The python-cryptography project is a special case, as it mostly wraps OpenSSL.
However, it provides its own \gls{asn1} parsing modules written in the Rust programming language\footnote{\url{https://cryptography.io/en/latest/faq}}.
The selection is also limited by the applicability of the used libraries.
For instance, BoringSSL which is used by the Google Chrome browser does not provide any \gls{api} or \gls{abi} stability guarantees and its authors recommend that the library should not be used by third parties\footnote{\url{https://boringssl.googlesource.com/boringssl/}}.

A further example of libraries which were not considered by the authors are LibreSSL\footnote{\url{https://www.libressl.org/}} or \gls{nss}\footnote{\url{https://firefox-source-docs.mozilla.org/security/nss/index.html}}.
LibreSSL, was created as a fork\footnote{\url{https://www.openbsd.org/papers/eurobsdcon2014-libressl.html}} from OpenSSL after the (in)famous Heartbleed vulnerability~(CVE-2014-0160).
LibreSSL was intended to be a more secure drop-in replacement for OpenSSL.
Therefore, those two libraries are mutually exclusive on a computer system.
Apart from those technical difficulties, LibreSSL is only natively supported on the OpenBSD operating system.
LibreSSL's impact on the overall ecosystem is considered low, since OpenBSD has a relatively low market share.
There are projects, such as OPNSense (based on the larger FreeBSD project), which offered official ports using LibreSSL but recently discontinued the support due to incompatibility reasons\footnote{\url{https://opnsense.org/opnsense-22-7-released/}} and maintenance burden.

\gls{nss} is the \gls{tls} library which is embedded into the Firefox browser.
Due to its domain-specific use case and special design, the authors were not able to create an isolated test setup using only the X.509 parsing module of the \gls{nss} library.
A simple test setup was preferred in order to ensure understandable and reproducible tests.
Therefore, custom and complex workarounds using, e.g., multiple container environments with custom build setups for each library were avoided.

For each library, two test runs were performed.
The libraries were obtained from the nixpkgs repository\footnote{\url{https://github.com/NixOS/nixpkgs}}.
The first run used a snapshot from October 2023.
The second run used a snapshot from March 2024.
The version numbers of the libraries stem from the used nixpkgs snapshot.

\subsection{Test Data}

Since the focus of this study is not on fuzzing and \emph{generating} test data, appropriate test data was \emph{collected}.
During the research, multiple test data providers were contacted and evaluated and finally the following dataset with collected X.509 certificates provided by Censys~\cite{censys2015} was found to be suitable:

\begin{quote}
    The Censys Universal Internet Dataset provides accurate, up-to-date records about Internet-facing hosts: both information that applies to the host as a whole as well as information about the services running on the host.
\end{quote}

The used dataset in this research is from July 1, 2023.
The dataset is provided in the avro file format which showed significant performance problems.
Therefore, the data was converted to compressed compressed \gls{json} files using the \gls{zst} algorithm.
Interestingly enough, those files were smaller than the avro files; the converted dataset had a size of 11.6~TiB.
The dataset is separated into multiple files where each file consists of $\approx 150,000$ certificates, called a “batch”.
A suitable download and converter script is included in the additional material to this paper.
        
\subsection{Data Acquisition}

For each library, a small program was created that reads a line-separated list of X.509 certificates in base64 encoding from its standard input stream.
Each line is decoded and fed into the X.509 parser of the relevant library.
In this test, only the parsing functionality of the libraries is used.
Therefore, no certificate validation functions are used.
Metadata about the testrun, such as the fingerprint of the tested certificate, the test duration, or the error code, is collected and stored in an SQLite database.

\begin{figure}[htb]
\centering
\includegraphics[width=0.45\textwidth]{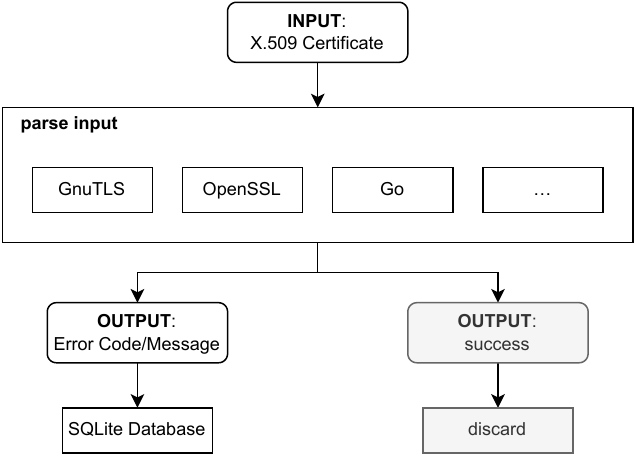}
\caption{Schematic figure of the test setup.}
\label{fig:schematic}
\end{figure}

\autoref{fig:schematic} shows the basic idea of the test setup.
The same certificates are fed to several different X.509 certificate parsers and the results are stored in a database.
Certificates which are parsed successfully are not stored in order to keep the database size minimal.
\autoref{tab:methods} shows the used library functions to parse and load the tested certificates.

\small

\begin{table}[htb]
    \centering
    \caption{The used library functions for parsing the tested X.509 certificates.}
    \begin{tabular}{ll}
        \toprule
        Library & Parsing Function \\
        \midrule
        GnuTLS & \texttt{gnutls\_x509\_crt\_import()} \\
  \hdashline
        Mbed~TLS & \texttt{mbedtls\_x509\_crt\_parse()} \\
  \hdashline
        OpenSSL & \texttt{d2i\_X509()} \\
  \hdashline
        wolfSSL & \texttt{wolfSSL\_X509\_d2i()} \\
  \hdashline
        cryptography & \texttt{cryptography.x509.load\_der\_x509\_certificate()} \\ 
  \hdashline
        Go stdlib & \texttt{x509.ParseCertificate()} \\ 
        \bottomrule
    \end{tabular}
    \label{tab:methods}
\end{table}
\normalsize

Due to the huge size of the dataset, a few optimizations were applied to the test setup.
In order to keep it reproducible and understandable, own isolated test programs were created for each tested library.
As a first step, all these test programs were run in parallel using the GNU parallel utility \cite{gnuparallel}.
Since SQLite was used as a database system, concurrent test runs had a poor performance, as the database library synchronized disk writes.
That architecture significantly increased the duration of the test suite.
Therefore, the test results were temporarily stored in compressed \gls{json} files and afterwards bulk imported into the SQLite database.
Each entry in the SQLite database contains a reference to the input file and the fingerprint of the parsed certificate.
In other words, each test can be mapped to the tested certificate.
After finishing the tests, the SQLite database contains 2,238,406,367 entries.

\section{Error Categories}

In order to create comparable results, the observed error messages were categorized into different groups.
After careful inspection of all unique error messages in the dataset, the authors have defined the following error categories.
For each library and each error category all the observed error strings were evaluated.
The authors manually assigned the appropriate error category.

\subsection{ASN1\_PARSE\_ERROR}

This error describes a lower level X.509 parsing error at the \gls{asn1} level.
The \gls{asn1} standard is used for the binary encoding of certificates.
An example for such an error is data that does not conform to the \gls{der} specification.
\exampleError{x509: malformed UTCTime}.

\subsection{CRYPTO\_UNSUPPORTED}

This error is related to unsupported cryptographic algorithms.
X.509 certificates require cryptographic algorithms for specific calculations, such as the validation check of the trust chain.
An example for such an error is a specified cryptographic algorithm that is not supported by the tested library.
\exampleError{x509: unsupported elliptic curve}.

\subsection{CRYPTO\_VALUE\_ERROR}

This error is related to invalid cryptograpic parameters.
Several algorithms have parameters with a direct impact to on their security, for instance, the public exponent of RSA.
An example of such an error is an insecure parameter in a supported cryptographic algorithm.
\exampleError{x509: invalid RSA modulus}.

\subsection{UNCATEGORIZED}

This is a special error category which is used for error messages without contextual meaning.
An example of such an error is “ok” which occurred in tests of the wolfSSL library.

\subsection{X509\_PARSE\_ERROR}

This error describes a higher-level X.509 parsing error.
Apart from parsing the underlying \gls{asn1} data, decoding a certificate might involve consecutive parsing steps.
An example for such an error is an invalid \gls{url}.
First, the raw string needs to be decoded from the \gls{asn1} raw data.
Subsequently, the decoded string is passed to a higher level \gls{url} parser.
\exampleError{x509: cannot parse URI}.

\subsection{X509\_UNSUPPORTED}

This error describes an unsupported X.509 value.
X.509 certificates use so called \glspl{oid} to describe specific functionality, such as permission flags.
An example for such an error is an \gls{oid} which is not supported by the tested library.
\exampleError{X509 - Unavailable feature, e.g. RSA hashing/encryption combination}.

\subsection{X509\_VALUE\_ERROR}

This error describes an invalid X.509 value.
In other words, the certificate could be parsed correctly but there are invalid values at a logical level.
An example for such an error is an invalid certificate version number or an unsupported date.
\exampleError{X509 - Signature algorithms do not match.}.

\section{Evaluation}

% \begin{itemize}
%     \item \todo{auswertungen wie in frankencert paper: most common issuer, etc.}
%     \item \todo{frankencert table 5 anschaun und analysen davon ableiten}
%     \item \todo{frankencert paper: Other checks, Seite 126 }
% \end{itemize}

Three parameters of the tested certificates were evaluated: Parsing Performance, Error Rates, and Error Categories.
Finally, noticeable errors were analyzed manually.
The following evaluation is based on the SQLite database which was created by the conducted test runs.

\subsection{Performance}

The parsing performance of the tested libraries was evaluated per batch. 
One batch file from the used dataset consists of $\approx 150,000$ different certificates.
The batch files were created to have a similar file size. Thus, the exact number of certificates per batch file can vary.
During the test runs, 2,000 batch files were tested.
All these certificates were parsed in sequential order and the duration was recorded.

\begin{figure}[htb]
\centering
\includegraphics[width=0.45\textwidth]{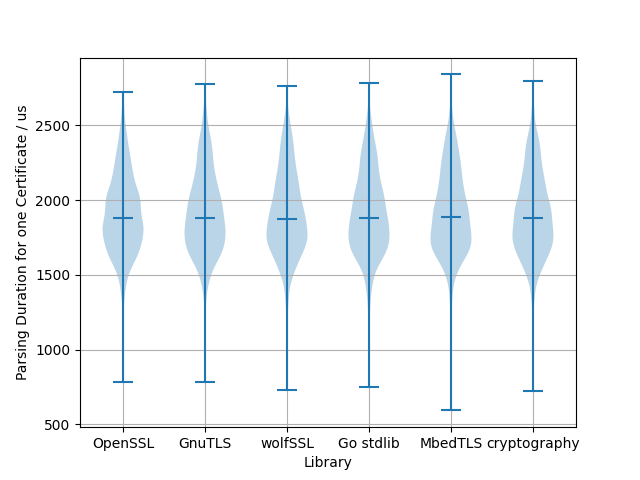}
\caption{A violin plot of the normalized parsing duration for one certificate file from the chosen dataset. One batch file contains between $\approx 70,000$ and $\approx 150,000$ certificates. 2,000 batch files were processed for this plot. The parsing duration of one certificate was calculated by dividing the parsing duration of a batch file by the number of contained certificates.}
\label{fig:chart-performance}
\end{figure}

\autoref{fig:chart-performance} shows a violin plot of the parsing duration.
This violin plot indicates the whole value range, the median, and the probability density of the data at different values.
Since the batch files are limited to a maximum file size, the number of contained certificates differs.
\autoref{fig:chart-performance} contains normalized numbers.
Comparing the median value, all libraries have a similar median parsing duration per batch file.
The wolfSSL and Mbed~TLS libraries have a fluctuation to lower values (= faster).

\subsection{Error Rates}

\begin{table}[htb]
    \centering
    \caption{Overview of the error rates among all tested libraries. The error rate is defined as the number of certificates that failed to parse divided by the number of tested certificates, i.e., $N = 186,576,846$.}
    \begin{tabular}{lrr}
    \toprule
Library & $n_e$ & $r_e = \frac{n_e}{N}$ \\
    \midrule
wolfSSL & 24,196,189 & 13.16\%\\
\hdashline
Mbed~TLS & 24,562,224 &  12.97\%\\
\hdashline
Go stdlib & 51,181 & 0.03\%\\
\hdashline
GnuTLS & 28,875 & 0.02\%\\
\hdashline
cryptography &21,492 & 0.01\%\\
\hdashline
OpenSSL & 4,803 & 0.003\%\\
    \bottomrule
    \end{tabular}
    \label{tab:rates}
\end{table}

\autoref{tab:rates} shows the error rates for all tested libraries.
The individual error rate $r_e$ was calculated for each library as follows: the number of errors $n_e$ divided by the number of parsed certificates $N$:

\begin{equation}
    r_e = \frac{n_e}{N}
\end{equation}

There are two libraries (wolfSSL and Mbed~TLS) with similar error rates (13.16\% and 12.97\%).
The other libraries all have error rates of $r_e \leq 0.03\%$.
Both wolfSSL and Mbed~TLS show fluctuations in the parsing duration and at the same time have the highest error rate.

\begin{table*}
    \centering
    \caption{Overview of the conducted tests and the number of occurrences of errors and error category. Each library was used to parse the same 186,576,846 certificates from the chosen dataset. The symbol “n.a.” means that no error messages belonging to that category were observed and therefore not considered by the tooling.}
\begin{tabular}{llrrrrrrrrr}
  \toprule
Lang. & Library & Version & \rot{\shortstack{ASN1\_PARSE\\ \_ERROR}} & \rot{\shortstack{CRYPTO\\ \_UNSUPPORTED}} & \rot{\shortstack{CRYPTO\_VALUE\\ \_ERROR}} & \rot{UNCATEGORIZED} & \rot{\shortstack{X509\_PARSE\\ \_ERROR}} & \rot{\shortstack{X509\\ \_UNSUPPORTED}} & \rot{\shortstack{X509\_VALUE\\ \_ERROR}} & \rot{\textbf{Sum}} \\
\midrule
C & GnuTLS & 3.8.1 & 6,321 & n.a. & n.a. & n.a. & 12,599 & n.a. & 9,955 & \textbf{28,875}  \\
  \hdashline
 &  & 3.8.3 & 6,321 & n.a. & n.a. & n.a. & 12,599 & n.a. & 9,955 & \textbf{28,875}\\
\midrule
 & Mbed~TLS & 3.4.1 & 84,397 & 183 & 1,561 & n.a. & 17,192 & 8,553 & 24,085,344 & \textbf{24,197,230}\\
  \hdashline
 & & 3.5.2 & 84,411 & 183 & 1,561 & n.a. & 17,227 & 8,553 & 24,092,409 & \textbf{24,204,344} \\
    \midrule
 & OpenSSL & 3.0.10 & 4,803 & n.a. & n.a. & n.a. & n.a. & n.a. & n.a. & \textbf{4,803}\\
  \hdashline
 & & 3.0.13& 4,803 & n.a. & n.a. & n.a. & n.a. & n.a. & n.a. & \textbf{4,803}\\
    \midrule
 & wolfSSL & 5.5.4 & 114,291 & 23 & n.a. & 11,426 & n.a. & n.a. & 24,070,449 & \textbf{24,196,189}\\
  \hdashline
 & & 5.7.0 & 467,257 & 23 & n.a. & 41 & n.a. & n.a. & 24,094,903 & \textbf{24,562,224}\\
    \midrule
    Go & stdlib & 1.20.7 & 40,736 & 46 & 374 & n.a. & 5,148 & n.a. & 4,877 & \textbf{51,181}\\
  \hdashline
    & &  1.22.1 & 40,736 & 46 & 374 & n.a. & 5,148 & n.a. & 4,877 & \textbf{51,181} \\
    \midrule
Python & cryptography &  41.0.4 & 19,982 & n.a. & n.a. & n.a. & n.a. & n.a. & 1,510 & \textbf{21,492}\\
  \hdashline
 v3.11& & 42.0.5 & 19,982 & n.a. & n.a. & n.a. & n.a. & n.a. & 1,510 & \textbf{21,492} \\
    \midrule
\end{tabular}
    \label{tab:cert-measures}
\end{table*}

\subsection{Error Categories}

\begin{figure}[htb]
\centering
\includegraphics[width=0.485\textwidth]{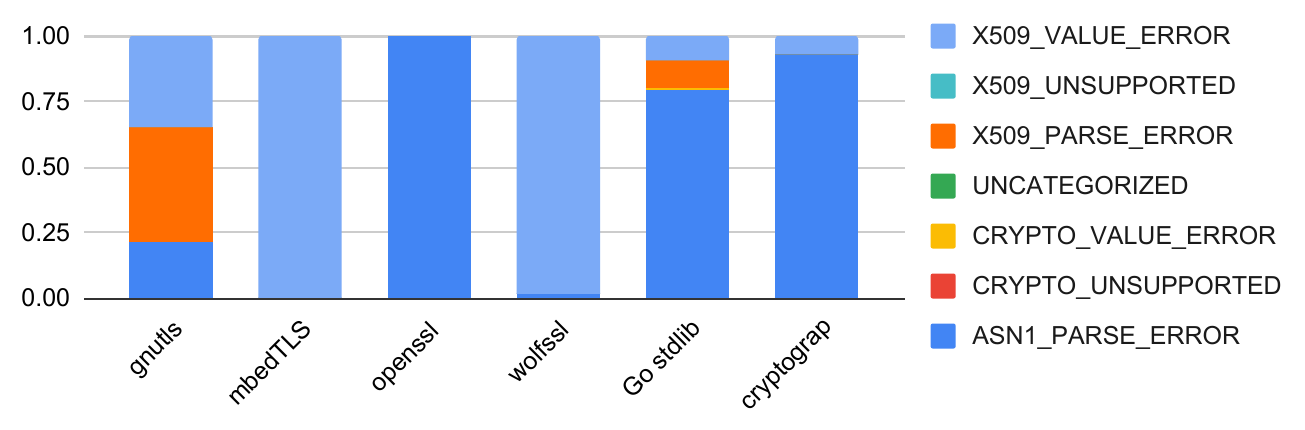}
\caption{Normalized distribution of error categories among the tested libraries.}
\label{fig:chart-categories}
\end{figure}

\autoref{fig:chart-categories} shows the normalized distribution of the error categories among the tested libraries.
The error categories were assigned by a manual approach and show some insights into the libraries' feature set.
GnuTLS and the Go stdlib have the highest amount of X509\_PARSE\_ERROR.
WolfSSL and Mbed~TLS have a very large amount of X509\_VALUE\_ERROR.
This could be an indication of missing features of those parsers.

% From frankencert paper:
% 
% \begin{itemize}
%     \item Certificate not yet valid
%     \item Certificate expired in its timezone
%     \item Server certificate not authorized for signing other certificates
%     \item Server certificate not authorized for server authentication
% \end{itemize}

\autoref{tab:cert-measures} shows the full results with the assigned error categories.
The GnuTLS library is relatively sparse in its error messages.
Only ASN1\_PARSE\_ERROR and X509\_PARSE\_ERROR could be mapped to GnuTLS' error strings.
Mbed~TLS and wolfSSL have a very similar number of X509\_VALUE\_ERRORs, thus the authors assume that those libraries could have some missing features in common.
Further analysis showed that $\approx 98\%$ of the certificates reported by both Mbed~TLS and wolfSSL as X509\_VALUE\_ERROR match.

In contrast to wolfSSL, Mbed~TLS reports errors about unsupported cryptographic algorithms or unsupported X.509 features.
Further, wolfSSL seems to have added previously missing error handling, since UNCATEGORIZED errors were reduced to 41 in version 5.7.0.
OpenSSL does not perform validations or checks above the \gls{asn1} level, since only ASN1\_PARSE\_ERRORs appear.
The Go standard library has fine grained error messages and most categories could be mapped to error strings.
Similar to OpenSSL, the python-cryptography library focuses on \gls{asn1}\_PARSE\_ERRORs.
Interestingly enough, the rewritten (in Rust) \gls{asn1} parser seems to be more restrictive as the C written parser in OpenSSL.
The lookup table of observed error strings and assigned error categories will be published in the supplemental material of this paper.

\begin{table*}[htb]
    \caption{Evaluation Results of Selected Errors. The Go stdlib has been used as a reference to find unique certificates that match a specific error condition. The chosen certificates were looked up in the database and the counts of all libraries were compared. Rows marked in yellow color contain discrepancies.}
    
    \begin{subtable}[h]{0.3\textwidth}
    \centering
    \caption{Number of unique certificates that were classified with “X509: invalid version”.}
    \begin{tabular}{lll}
        \toprule
        Library & Version & Count  \\
        \midrule
\rowcolor{yellow} GnuTLS & 3.8.1 & 0 \\
  \hdashline
Mbed~TLS & 3.5.2 & 160 \\
  \hdashline
\rowcolor{yellow} OpenSSL & 3.0.13 & 0 \\
  \hdashline
wolfSSL & 5.7.0 & 160 \\
  \hdashline
Go stdlib & 1.22.1 & 160 \\
  \hdashline
cryptography & 42.0.5 & 160 \\
        \bottomrule
    \end{tabular}
    \label{tab:x509-version}
    \end{subtable}
\hfill
\begin{subtable}[h]{0.3\textwidth}
    \centering
    \caption{Number of unique certificates that were classified with “x509: invalid RSA public exponent”.}
    \begin{tabular}{lll}
        \toprule
        Library & Version & Count  \\
        \midrule
\rowcolor{yellow} GnuTLS & 3.8.1 & 3 \\
  \hdashline
\rowcolor{yellow} Mbed~TLS & 3.5.2 & 4 \\
  \hdashline
\rowcolor{yellow} OpenSSL & 3.0.13 & 3 \\
  \hdashline
wolfSSL & 5.7.0 & 264 \\
  \hdashline
Go stdlib & 1.22.1 & 264 \\
  \hdashline
\rowcolor{yellow} cryptography & 42.0.5 & 3 \\
        \bottomrule
    \end{tabular}
    \label{tab:x509-rsa}
    \end{subtable}
\hfill
\begin{subtable}[h]{0.3\textwidth}
    \centering
    \caption{Number of unique certificates that were classified with “x509: invalid ECDSA parameter”.}
    \begin{tabular}{lll}
        \toprule
        Library & Version & Count  \\
        \midrule
\rowcolor{yellow} GnuTLS & 3.8.1 & 0 \\
  \hdashline
Mbed~TLS & 3.5.2 & 17 \\
  \hdashline
\rowcolor{yellow} OpenSSL & 3.0.13 & 0 \\
  \hdashline
wolfSSL & \textbf{5.5.4} & 17 \\
  \hdashline
\rowcolor{yellow} wolfSSL & \textbf{5.7.0} & 3 \\
  \hdashline
Go stdlib & 1.22.1 & 17 \\
  \hdashline
\rowcolor{yellow} cryptography & 42.0.5 & 0 \\
        \bottomrule
    \end{tabular}
    \label{tab:x509-ecdsa}
    \end{subtable}

\end{table*}

\subsection{Selected Errors}

Four concrete error messages which are easy to understand were picked from the Go stdlib list.
The sha256 fingerprints of the certificates showing these parsing errors were looked up in the results database.
Subsequently, the test results of these certificates were queried for all tested libraries and the number of matches were counted appropriately.

\subsubsection{x509: invalid version}

There are certificates in the dataset which use invalid X.509 version numbers.
According to RFC~5280~\cite{rfc5280} the version MUST be 2 or 3.
However, implementations are encouraged to prepare support for every version number.
\autoref{tab:x509-version} shows the results of this evaluation; 160 problematic certificates were tested.
GnuTLS and OpenSSL do not implement checks for the version number.
Mbed~TLS, wolfSSL, the Go stdlib, and python-cryptography have checks in place that reject certificates with version numbers different from 2 or 3.

\subsubsection{x509: invalid RSA public exponent}

There are certificates in the dataset which have invalid RSA public exponents.
RFC~2313~\cite{rfc2313} states that the public exponent used for RSA encryption may be standardized for specific applications.
The value $F_4 = 65537$, known as a Fermat number, is usually used for the public exponent in RSA encryption.
As shown in \autoref{tab:x509-rsa}, the Go stdlib and wolfSSL fail to parse the same 264 certificates due to invalid RSA exponents.
This is remarkable, since wolfSSL and the Go stdlib have different code bases even in different programming languages. 

\subsubsection{x509: invalid ECDSA parameter}
\label{sec:x509-ecdsa}

RFC~6979~\cite{rfc6979} states that there are specific constraints for the used \gls{ecdsa} parameters.
As shown in \autoref{tab:x509-ecdsa}, the dataset contains at least 17 problematic certificates.
GnuTLS, OpenSSL, and python-cryptography do not implement checks for those parameters.
Mbed~TLS, wolfSSL, and the Go stdlib do check those requirements.
The tests revealed that wolfSSL regressed between the tested versions 5.5.4 and 5.7.0, as the number of rejected certificates dropped from 17 to 3.

The authors investigated the potential regression in wolfSSL.
There are inconsistencies in the library according to the build flags used in the \texttt{configure} script.
When wolfSSL is compiled with the default build settings, then all 17 affected certificates are rejected in both tested versions.
Using the same settings as used in nixpkgs, including the \texttt{-{}-enable-all} flag, then only 3 certificates are rejected by wolfSSL 5.7.0; wolfSSL 5.5.4 rejects all 17 in both cases.
Due to the authors' lack of knowledge of wolfSSL's inner workings and its build toolchain, it was not possible to pinpoint this discrepancy's root cause.
It is worth mentioning that the ZLint~\cite{kumar2018}\footnote{\url{https://github.com/zmap/zlint}} tool fails to parse all 17 affected certificates, too.
ZLint is a linting tool for X.509 certificates which is widely used in research to check the consistency of certificates with applicable standards.

\subsubsection{x509: cannot parse URI}

This error is specific for the Go stdlib.
The dataset contains 5,025 certificates that cannot be parsed by the Go stdlib.
Among others, Go rejects certificates which are issued for \texttt{*.local} domains.
The other tested libraries do not seem to implement checks for \gls{uri} conformity, since no other errors were observed.

\section{Discussion}

This study used a mixed approach for evaluating the implemented measures.
The error strings and related code were examined manually and an appropriate entry was added to a lookup table.
Subsequently, the evaluation was automated via multiple Python scripts or spreadsheet formulas.

Such an approach is sensitive to human error; especially because the context of the error message is not 100 percent clear in all cases.
For instance, GnuTLS provides error messages in the following format:

\begin{itemize}
    \item Error in the certificate
    \item ASN1 parser: Error in DER parsing
    \item Duplicate extension in X.509 certificate.
\end{itemize}

Compared to error messages from, e.g., python-cryptography, GnuTLS' error messages are relatively sparse.
The Python library instead provides detailed information about the error category and the cause of the error, e.g.,

\begin{quote}
error parsing asn1 value: ParseError \{ kind: UnexpectedTag \{ actual: Tag \{ value: 20, constructed: true, class: Application \} \} \}.
\end{quote}
However, most errors from the Python library are such ParseErrors.
Hence, the considerably better error messages do not contribute to this study.

The quality of the results relies on high-quality error messages.
Most libraries suffer from poor error reporting when parsing X.509 certificates. 
The Go standard library and Mbed~TLS provide the most versatile error reporting among the examined libraries.
Both libraries have understandable error messages and cover most error categories defined in this study.
In contrast to that, OpenSSL has error messages which are often hard to understand, such as “error:0680009B:asn1 encoding routines::too long”.
However, there are error codes included in the error message which could be used to search for the root cause in the code base.

Based on the observed error messages, the authors conclude that some libraries perform additional checks and reject certain X.509 certificates.
A good example is the Go standard library which performs additional checks that were not observed by other libraries.
For instance, the Go standard library rejects certificates with malformed IP addresses or invalid cryptographic parameters.
The wolfSSL library seems to be in active development, since UNCATEGORIZED, TODO-like error codes, e.g., “ok” or “bad function argument”, were fixed between the tested versions.
A further indicator for this hypothesis is a regression which was found in wolfSSL's certificate parser, cf. Section~\ref{sec:x509-ecdsa}.
During the course of writing this paper, the authors were not able to pinpoint the root cause of this regression.
The certificates causing this discrepancy need to be analyzed in order to identify the relevant parts for the observed errors.
The usage of further software testing methods, e.g., test runs with enabled memory sanitizers or special debug builds, are possible future research.

The libraries with the highest error rates (cf. \autoref{tab:rates}) are Mbed~TLS and wolfSSL.
Both libraries have a very similar error rate of $\approx 13\%$ and the affected certificates match by a significant percentage.
These two libraries most likely share code or at least share design decisions of their parsing modules.
The authors conclude that Mbed~TLS and wolfSSL have incomplete X.509 parsers which do not handle all required corner cases.
An indicator for this hypothesis is that Mbed~TLS and wolfSSL fail to parse the same certificates in the X509\_VALUE\_ERROR category.
OpenSSL seems to only perform plain \gls{asn1} parsing with no additional checks, as it has the lowest error rate of all examined libraries and only ASN1\_PARSE\_ERRORs were observed.

Despite the fact that the ecosystem is fragmented by the existence of many different X.509 implementations which do not yield the same results, an open question remains: “Do these differences in conformity have an impact on the operational security of X.509 or \gls{asn1}?”
A future study examining more \gls{tls} libraries, the reasons for the observed parsing errors, and possible obstacles in the interoperability might be worthwhile.
Examining the possibility of accepting invalid certificates as valid might be of peculiar interest.

\section{Conclusion}

Due to the internet's heterogeneous technology landscape, it is crucial that all involved parties provide consistent results when parsing potentially untrusted data.
This research paper revealed that the tested X.509 parsing libraries \emph{do not} provide consistent parsing results.
For instance, wolfSSL and Mbed~TLS both fail to parse $\approx 13\%$ of the tested certificates while OpenSSL only fails to parse $0.003\%$.
Since this situation needs to be improved, this paper introduces an approach for testing X.509 parsing modules by classifying their errors.
The authors expect that unified parsing errors across multiple implementations will improve the traceability of errors which are otherwise hard to detect.

The authors would answer the formulated research question \resQuestion with a distinct \emph{yes}.
The involved standards are complex and contain a lot of legacy.
Moreover, they are relatively large and contain certain optional features.
Therefore, as also shown in Tatschner et al. \cite{tatschner2023ares}, the degree of autonomy for implementations is high and multiple different design choices are possible without violating the standards.

Due to the worldwide deployment of X.509-based technologies, it is not possible to simply replace these technologies with more modern ones.
Therefore, it is crucial to improve the current state of the art as good as possible.
Creating and maintaining automated tests verifying the conformity of libraries as much as possible might help to improve the ecosystem.

% https://www.bsi.bund.de/SharedDocs/Downloads/DE/BSI/Publikationen/TechnischeRichtlinien/TR02103/BSI-TR-02103.pdf?__blob=publicationFile&v=2

\begin{anonsuppress}
\section{Data Availability}

Supplementary material is provided under the CC BY 4.0 Legal Code license at \url{https://rumpelsepp.org/projects/parseval}.
Among others, the material provides the raw data of this study.

\begin{acks}
This work was founded by the German Federal Ministry for Economic Affairs and Climate Action~(BMWK) under Grant Number~13I40V010A, project “PoQsiKom”.
\end{acks}
\end{anonsuppress}

%%
%% The next two lines define the bibliography style to be used, and
%% the bibliography file.
%\bibliographystyle{ACM-Reference-Format}
%\bibliography{bibliography}

% Must start on new page:
% https://www.ares-conference.eu/submission-guidelines
\clearpage

\printbibliography

\end{document}